\documentstyle[amsfonts]{JHEP}
\title{Strongly Coupled CFT in FRW Universe from AdS/CFT Correspondence}
\author{K. Koyama \\  Graduate School of Human and Environment 
Studies, Kyoto University, Kyoto  606-8501, Japan
\\ \email{kazuya@phys.h.kyoto-u.ac.jp}}
\author{J. Soda
\\ Department of Fundamental Sciences, FIHS, Kyoto University,
       Kyoto, 606-8501, Japan
\\ \email{jiro@phys.h.kyoto-u.ac.jp}}

\abstract{
We develop a formalism to calculate the effective action of 
the strongly coupled conformal field theory (CFT) 
in curved spacetime. The effective action of the CFT is
obtained from AdS/CFT correspondence. 
The anti de-Sitter (AdS) spacetime has various slicing 
which give various curved spacetime on its boundary. 
We show the de Sitter
spacetime and the Friedmann-Robertson-Walker (FRW) universe 
can be embedded in the AdS spacetime and derive the
scalar two-point function of the conformal fields in those 
spacetime. In curved spacetime, the two-point function depends
on the vacuum state of the CFT. A method to specify 
the vacuum state in AdS/CFT calculations is shown. 
Because the classical action in AdS spacetime diverges
near the boundary, we need the counter terms to regulate the
result. The simple derivation of the counter terms 
using the Hamilton-Jacobi equation is also presented in the appendix.
}
\preprint{
\begin{tabular}{l}
hep-th/0101164
\end{tabular}}
\begin{document}
\section{Introduction}
There are many matter fields in the Standard Model of particle
physics. Supersymmetric theory or string theory predicts
much larger number of matter fields.
In the early universe, these matter fields are effectively
massless and then  behave as the conformaly invariant fields.
Quantum effects of them can play an important role
in the history of the universe.
One of the examples is the trace anomaly of the energy-momentum
tensor of the matter fields. The trace anomaly
acts like a cosmological constant and could lead to the
inflationary universe \cite{M1,H2}.

In order to know the quantum effects of the matter fields, 
we should know the effective action which describes
the effects of the matter fields. 
A way of calculating
the effective action is to use AdS/CFT
correspondence \cite{H2,A/C}.
AdS/CFT correspondence states
quantum theory of CFT on $d$-dimensional boundary
of the Anti-de Sitter (AdS) spacetime can be described by the
classical theory in the $(d+1)$--dimensional AdS bulk.
This correspondence is on the ground of
duality in the string theory. The classical limit of type IIB
superstring theory (supergravity)
on five dimensional AdS spacetime times a
five sphere ($AdS_5 \times S^{5}$)
is dual to ${\cal N}=4$ supersymmetric $SU(N)$ Yangs-
Mills (YM) theory for large $N$. The CFT parameters are related to
the supergravity parameters by $l^3/G=2 N^2/\pi$, where
$G$ is the 5-dimensional Newton constant and $l$ is the
curvature radius of $AdS_5$. The curvature scale of the
$AdS_5$ spacetime $l$ is given by $l^2=\sqrt{g^2_{YM} N} \: l_s^2$
where $g_{YM}$ is the coupling constant of YM theory and
$l_s$ is the string length. In order to trust the classical
theory in the bulk we need $(l/l_s)^4=g^2_{YM} N \gg 1$. Thus, the
dual CFT should have large N.

Practically, the AdS/CFT correspondence gives an efficient way of
calculating a non-local effective action in the boundary spacetime.
Let us consider the gravitational theory.
In the dual theory on the boundary, gravity is coupled to the
matter fields. By performing the path integral over the matter
fields, an effective action that is the functional of the
background metric is obtained. In general, the effective
action contains non-local terms which make the calculations
difficult. If one uses the AdS/CFT correspondence,
the non-local effective action can be found from the 
classical action in the $AdS_{d+1}$ spacetime.
Indeed, quantum corrections of the matter fields on the graviton
propagator were calculated using an effective action derived
from AdS/CFT correspondence and the result was shown to be agreed
with the old calculations done in $d$-dimensional
quantum field theory \cite{EA}.
So far much of the calculations have
been done in the flat spacetime. In order to observe the effects
of the quantum matter in the early universe, we should know the
effective action in curved spacetime. Several authors extended 
the calculations to the de Sitter spacetime \cite{H2,H1,No,N}. 
We will extend their works to several directions.

The main claim of our paper
is that the effective action in curved spacetime can be constructed
by choosing appropriate slicing in the AdS bulk. 
The AdS spacetime has various slicing 
which give various curved spacetime on its boundary. For example, 
the de Sitter spacetime and the 
Friedmann-Robertson-Walker (FRW) universe
can be embedded into the AdS spacetime. 
We will explicitly derive the two-point 
function of the conformal field in the de Sitter spacetime and the 
FRW spacetime.
In curved spacetime, the two-point function of the quantum
field depends on the choice of the vacuum. A method to 
specify the vacuum state of the CFT in AdS/CFT calculations
will be presented.

We concentrate our attention on the scalar field theory
in the bulk. In the dual theory on the boundary,
there is a scalar field $\phi$ coupled with conformal 
matter fields $\psi$, say
\begin{equation}
S[\phi,\psi]=\int d^d x \sqrt{-h}
\left(- 
\partial_{\mu} \phi
\partial^{\mu} \phi +e^{-2 \phi} {\cal L}_{matter}[\psi]
\right),
\end{equation}
where ${\cal L}_{matter}$ is the Lagrangian for conformal
matter fields $\psi$ 
and $h$ is the metric of the $d$-dimensional spacetime.
As in the gravitational theory,
the generating functional can be obtained by
performing the path integral over the matter fields $\psi$ 
to be a functional of the scalar field
\begin{equation}
W[\phi]= \int {\cal D} \psi e^{-i S[\phi,\psi]}.
\end{equation}
AdS/CFT correspondence
gives it as the generating functional of the dual
scalar operator ${\cal O}$
coupled with the scalar field $\phi$ on the boundary
\begin{eqnarray}
&&W[\phi] = W_{CFT}=
\left \langle \exp \left(i \int d^d x \sqrt{-h} 
\phi {\cal O} 
\right)
\right \rangle \nonumber\\
&=& \exp \left(i \int d^d x \sqrt{-h} \phi 
\langle {\cal O} \rangle -\frac{1}{2} \int
d^d x d^d x' \sqrt{-h} \sqrt{-h}
\phi(x)\phi(x') \langle {\cal O}(x) {\cal O}(x')
\rangle + \cdot \cdot \cdot \right). \nonumber\\
\label{1-1}
\end{eqnarray}
Varying the generating functional $n$ times with respect to
$\phi$ gives the $n$-point function of operator ${\cal O}$. 

The outline of our paper is as follows. In section 2, we review the
method to calculate the effective action in flat spacetime 
using the invariant Green function in AdS spacetime.
In section 3, we extend the calculation to the de Sitter spacetime.
We will explain how the information
of the vacuum state of the boundary theory is encoded 
in the AdS/CFT calculation.
Then the nature of the CFT in de Sitter spacetime is presented.
Section 4 is devoted to the formulation for the FRW universe.
In the appendix, the derivation of the counter terms via 
Hamiltonian-Jacobi equation is shown, 
which are needed to regulate the action on the boundary. 

%%%%%%%%%%%%%%%%%%%%%%%%%%%%%%%%%%%%%%%%%%%%%%%%%%%%%%%%%%%%%%%%%%%%%%%%
%%%%%%%%%%%%%%%%%%%%%%%%%%%%%%%%%%%%%%%%%%%%%%%%%%%%%%%%%%%%%%%%%%%%%%%%

\section{A review}

AdS/CFT correspondence states
the generating function of the CFT $W_{CFT}$
in $d$-dimensional spacetime
can be obtained by the classical theory in the
$d+1$-dimensional AdS spacetime $AdS_{d+1}$ \cite{A/C}. 
The correspondence is written as
\begin{equation}
W_{CFT}= \left \langle \exp \left(i \int_{\cal B} \phi_B {\cal O} 
\right)
\right \rangle =e^{i I[\phi_B]},
\label{2-1}
\end{equation}
where ${\cal O}$ is the operator of the CFT coupled with
the scalar field $\phi_B$ in the d-dimensional boundary 
${\cal B}$ of the $AdS_{d+1}$.
$I[\phi_B]$ is obtained from $d+1$-dimensional action of the 
scalar field $\phi$
in $AdS_{d+1}$ 
\begin{equation}
S[\phi]= -\frac{1}{2}\int d^{d+1} X \sqrt{-g}
\left(\partial_A \phi \partial^A \phi + m^2 \phi^2 \right),
\label{2-2}
\end{equation}
by inserting the solution of the classical
field equation which approaches $\phi_B$ on the boundary.
For simplicity, we concentrate our attention on the
generating function of the two-point function.
The extension to $n$-point function can be easily carried out
\cite{F}. The calculation of $I[\phi_B]$ is done as follows.
First we should find the solution of the field equation in
$AdS_{d+1}$ spacetime which approaches $\phi_B$ on the boundary
${\cal B}$. The solution can be constructed from the
Dirichlet Green function, which satisfies
\begin{eqnarray}
(\Box_{d+1} -m^2)G(X;X') &=& -\frac{\delta^{d+1}(X-X')}
{\sqrt{-g}}, \nonumber\\
G(X;X') \vert_{ {\cal B}} &=& 0,
\label{2-3}
\end{eqnarray}
where $\Box_{d+1}$ is the Laplacian operator 
in the $AdS_{d+1}$ spacetime and $X$ is the coordinate of 
$AdS_{d+1}$.

From the general formalism of the Green function, the
solution satisfying the field equation and boundary condition
$\phi \vert_{{\cal B}} =\phi_B$ can be written as
\begin{equation}
\phi(X) = -\int_{{\cal B}} d^d x' \sqrt{-\gamma}
G_{\cal B} (X;x') \phi_B(x') ,\quad
G_{\cal B}=
\left(n^{A} \frac{\partial}{\partial X'^{A}} G(X;X')
\right)_{{\cal B}},
\label{2-4}
\end{equation}
where $\gamma$ is the induced metric on the boundary ${\cal B}$, 
$n^{A}$ is the unit vector normal to ${\cal B}$ and
$x$ is the coordinate of the boundary.
$G_{\cal B}(X;x')$ is often called bulk-boundary propagator.
Our task is to find the Dirichlet Green function in $AdS_{d+1}$.
One way is to use the invariant Green function 
that depends only on the geodesic distance of the 
AdS spacetime \cite{IG}. The invariant Green function
is defined as
\begin{equation}
G(X,X')= G(\mu(X,X')),
\end{equation}
where $\mu(X,X')$ is the geodesic distance between
$X$ and $X'$. For $X \neq X'$, $G(\mu)$ should satisfy 
the field equation
$(\Box_{d+1}-m^2)G(\mu)=0$ which can be rewritten 
in terms of $\mu$ as 
\begin{equation}
\frac{d^2 G(\mu)}{d \mu^2} + (d/l) \coth(\mu/l) 
\frac{d G(\mu)}{d \mu}  -m^2 G(\mu)=0,
\label{2-4-1}
\end{equation}
where $l$ is the curvature radius of the AdS spacetime.
After defining the new variable 
\begin{equation}
W = \frac{2}{\cosh (\mu/l) +1},
\end{equation}
the equation (\ref{2-4-1}) can be converted into a
hypergeometric equation. The equation (\ref{2-4-1})
has two independent solutions.
Among them, the solution
which satisfies the Dirichlet boundary condition at the boundary
$G(\mu) \vert_{\cal B}=0$ is given by
\begin{equation}
G(\mu)=G_0 l^{1-d} W^{\Delta}
 F \left(\Delta,\Delta-\frac{d}{2}+\frac{1}{2},2
\Delta -d+1,W \right) ,
\label{2-5}
\end{equation}
where $F$ is the hypergeometric function,
\begin{eqnarray}
G_0 &=& \frac{i \Gamma[\Delta]}{\pi^{d/2} 2^{2 \Delta+1}
\Gamma[\Delta-\frac{d}{2}+1]}, \nonumber\\
\Delta &=& \frac{1}{2}(d+\sqrt{d^2+4 m^2 l^2}).
\end{eqnarray}
The normalization $G_0$ is determined so as to have the same 
singularity with the Green function in flat spacetime 
for $\mu \to 0$
\begin{equation}
\lim_{\mu \to 0} G(\mu)= \frac{i \Gamma \left[\frac{d+1}{2}
\right]}{2(d-1) \pi^{(d+1)/2}} \mu^{1-d}.
\end{equation}
At the boundary of the AdS spacetime, 
$\mu \to \infty$ and $W \to 0$. The invariant Green function
near the boundary is given by
\begin{equation}
G(\mu) \to  G_0 W^{\Delta},
\label{2-7-0}
\end{equation}
which vanishes at the boundary as expected.  
Integrating by parts and using the equation of motion, 
the action $S[\phi]$ is rewritten as
\begin{equation}
S[\phi]= -\frac{1}{2} \int_{{\cal B}} d^d x
\sqrt{-\gamma} n^{A} \phi \partial_{A} \phi,
\label{2-7}
\end{equation}
where we have used the fact the Green function (\ref{2-5})
vanishes at the horizon and then surface term at the horizon 
does not contribute to the action.
Inserting the solution (\ref{2-4}) into the action (\ref{2-7}),
we find
\begin{eqnarray}
I[\phi_B] &=& \frac{1}{2} \int_{{\cal B}}
d^d x d^d x'\phi_B(x) \phi_B(x')  \nonumber\\
&\times&
\left(
 \sqrt{-\gamma(x)} \sqrt{-\gamma(x')} n^{A}(X) n^{B}(x')
 \frac{\partial^2}{\partial X^{A} \partial X'^{B}}G(X;X')
 \right)_{{\cal B}}.
\label{2-8}
\end{eqnarray}

So far, we do not specify the geometry of the boundary.
In AdS spacetime various spacetime can be embedded.
The most simple one is the Minkowski spacetime.
We use the Poincare coordinate to represent the
$AdS_{d+1}$ 
\begin{equation}
ds^2=\left(\frac{l}{z} \right)^2 (dz^2 -d \tau^2 +\delta_{ij}
dx^i dx^j).
\label{2-9}
\end{equation}
Each slicing of $z=const.$ and the boundary ${\cal B}$ are
Minkowski spacetime. Because the boundary is located at $z=0$,
we first consider the slicing $z=\epsilon$ and then take the
limit $\epsilon \to 0$.
The invariant $W$ can be expressed in terms of the
Poincare coordinate as
\begin{equation}
W=4 \frac{z z'}{(z+z')^2+\vert x-x' \vert^2},
\label{2-10}
\end{equation}
where $\vert x-x' \vert^2=-(\tau-\tau')^2+
\vert x^i-x^{'i} \vert^2$. Then from (\ref{2-7-0}), 
we can show 
\begin{equation}
\lim_{z=z'=\epsilon \to 0} \frac{\partial^2}
{\partial z \partial z'}
G(z,x;z'x')=G_0 l^{1-d} 2^{2 \Delta} \Delta^2
\epsilon^{2(\Delta-1)} \frac{1}
{\vert x-x' \vert^{2\Delta}}.
\end{equation}
Because $\sqrt{-\gamma}= 
(\epsilon/l)^{-d}$ and 
$n^{A} (\partial/\partial X^{A})= -(\epsilon/l)
\partial/\partial z$, 
we find
\begin{equation}
I[\phi_B] = \frac{1}{2} \int_{{\cal B}}
d^d x d^d x'
\phi_B(x) \phi_B(x') \epsilon^{2(\Delta-d)} l^{d-1} 
\left(
G_0 2^{2 \Delta} \Delta^2
\frac{1}{\vert x-x' \vert^{2\Delta}} \right).
\label{2-11}
\end{equation}
Then renormalizing $\phi_B$ as
\begin{equation} 
\tilde{\phi}_B = \frac{\Delta}{2\Delta-d}  \epsilon^{\Delta-d}
l^{(d-1)/2} \phi_B,
\label{2-12}
\end{equation}
the two point function of the CFT is obtained from
(\ref{1-1}) and (\ref{2-1}) as
\begin{eqnarray}
\langle {\cal O}(x) {\cal O}(x') \rangle &=& 
\frac{(-i)\delta^2 I}{\delta \tilde{\phi}_B(x) 
\delta \tilde{\phi}_B(x')} 
= C_2 \frac{1}{\vert x-x' \vert^{2 \Delta}}, 
\nonumber\\
C_2 &=& 2 
\left(\Delta-\frac{d}{2} \right) \frac{\Gamma[\Delta]}
{\pi^{d/2} \Gamma[\Delta-\frac{d}{2}]} .
\label{2-13}
\end{eqnarray}
The need of the factor $\Delta/(2 \Delta-d)$ comes from 
the limiting procedure. We first 
take the limit $\epsilon \to 0$ in constructing the
bulk-boundary propagator (\ref{2-4}). That is, we used the
Dirichlet Green function which vanishes at $z=0$ and not
at $z=\epsilon$. More detailed discussions are seen in
\cite{KW}

\section{CFT in de Sitter spacetime}
\subsection{CFT in de Sitter spacetime}
The AdS spacetime has various slicing which realize various
curved spacetime on it. These slicing can be obtained by the
coordinate transformation from the Poincare coordinate.
Let us consider the coordinate transformation
\begin{equation}
z = \eta \sinh y , \quad \tau =  \eta \cosh y.
\label{2-16}
\end{equation}
The metric (\ref{2-9}) becomes
\begin{equation}
ds^2= \left(\frac{l^2}{\sinh^2 y}\right)
\left(d y^2 + \frac{1}{\eta^2} (-d \eta^2
+ \delta_{ij} dx^i dx^j )\right).
\label{2-17}
\end{equation}
On the slicing $y=y_0$, the embedded spacetime 
is the de-Sitter spacetime with the Hubble parameter 
$H=l^{-1} \sinh y_0$.

If we use the invariant Green function in calculating the
bulk-boundary propagator, 
the calculation of the generating function is essentially
the same with the Minkowski spacetime.
We should merely write the invariant length in terms of 
the associated coordinate. 
One subtle point is the procedure of taking the
limit to the boundary because the de Sitter spacetime
with specific $H$ is realized on only one slicing $y=y_0$
in the AdS spacetime. We need to 
introduce the parameter $\epsilon$ which
controls the limit to the boundary, $z \propto \epsilon$.
We can define
\begin{equation}
\epsilon=  \sinh y_0.
\label{2-22}
\end{equation}
Note that because $\sinh y_0=H l$,
we should take $l \to 0$ to fix $H$ in taking the limit
$\epsilon \to 0$. Once the limiting procedure is established,
the effective action can be obtained
using (\ref{2-8}).
$W$ can be expressed in terms of the coordinate (\ref{2-17}) as
\begin{equation}
W = 4 \frac{ \eta \eta' \sinh y \sinh y'}{
-(\eta^2+\eta'^2-2 \eta \eta' \cosh(y+y'))+
\vert x^i-x^{'i} \vert^2}.
\end{equation}
Then we find
\begin{eqnarray}
\lim_{\epsilon \to 0} \frac{\partial^2}
{\partial y \partial y'} 
G(y,x;y'x')&=& G_0 2^{2 \Delta} \Delta^2
\epsilon^{2(\Delta-1)} 
\left(
\frac{ \eta \eta'}{-(\eta-\eta')^2+\vert x^i- x^{'i} \vert^2} 
\right)^{\Delta} \nonumber\\
&=& G_0  2^{2 \Delta} \Delta^2 \epsilon^{2(\Delta-1)}
\sigma^{-2\Delta},
\end{eqnarray}
where $\sigma$ is the invariant length of the
de Sitter spacetime. Denoting the metric of
the de Sitter spacetime as
\begin{equation}
h_{\mu \nu}dx^{\mu}dx^{\nu}=\frac{1}{\eta^2}(-d \eta^2
+\delta_{ij}dx^i dx^j),
\label{4-46-0}
\end{equation}
we have $\sqrt{-\gamma}
=(\epsilon/l)^{-d} \sqrt{-h}$.
Then (\ref{2-8}) becomes 
\begin{equation}
I[\phi_B] = \frac{1}{2} \int_{{\cal B}}
d^d x d^d x' \sqrt{-h(x)} \sqrt{-h(x')}
\phi_B(x) \phi_B(x') \epsilon^{2(\Delta-d)} l^{d-1} 
G_0 2^{2 \Delta} \Delta^2 \sigma^{-2 \Delta}.
\end{equation}
Renormalizong $\phi_B$ as in (\ref{2-12}), 
the two point function of the CFT in de Sitter
spacetime is obtained as
\begin{eqnarray}
\langle {\cal O}(x) {\cal O}(x') \rangle_{dS}
= C_2 \sigma^{-2\Delta}.
\label{4-46}
\end{eqnarray} 

\subsection{Vacuum state of CFT}
In the previous subsection, we obtained the generating functional
of the CFT from invariant Green
function of $AdS_{d+1}$ spacetime.
In curved spacetime, the correlation function of the CFT 
depends on the choice of the vacuum. In the AdS/CFT correspondence,
the vacuum state on the boundary theory corresponds to 
the boundary condition of the bulk-boundary 
propagator \cite{kr,Vak,S}.
Using the invariant Green function in constructing the
bulk-boundary propagator will specify the vacuum state of the CFT.

To see this fact, it is convenient to do the calculations
in momentum spacetime.
The bulk-boundary propagator in the momentum space can be
constructed from the solution of the field equation 
in $AdS_{d+1}$.
\begin{equation}
(\Box_{d+1} -m^2) \phi(y,x) = 0.
\label{3-32}
\end{equation}
In the coordinate (\ref{2-17}),
the solution can be separated with respect to
the coordinate of the bulk $y$ and the
d-dimensional boundary $x$,
\begin{equation}
\phi(y,x)= \int d p' \: f_{p'}(y) Y_{p'}(x).
\label{3-33}
\end{equation}
Here we introduced the scalar harmonics $Y_{p'}$ 
in the d-dimensional de Sitter spacetime.
We notice that there are two ambiguities of the boundary conditions
in the solution. One is the
boundary condition at the horizon of the $AdS_{d+1}$
spacetime which specifies $f_{p'}(y)$. Another is the choice
of the time slicing in the d-dimensional spacetime
which is determined by the choice of the harmonics
$Y_{p'}(x)$. The freedom of the choice in $f_{p'}(y)$ may
correspond to the freedom to choose different Lorentzian
propagators in the CFT (Feynman propagator, retarded propagator
etc.) \cite{kr}. The choice of the time slicing will lead to the
choice of the vacuum state in the CFT in curved spacetime.
So, choosing an appropriate harmonics, we can
find the CFT in a desired vacuum.

Now the question is which boundary condition
leads to the result (\ref{4-46}).
We will show taking the Euclidean boundary condition 
gives the result.
For computational simplicity,
we assume a closed de Sitter spacetime
and take $d=4$. Then, we can perform the calculations
in the Euclidean spacetime in which the spacetime is
four sphere $S^4$ \cite{H2,H1}.
We first solve the field equation in the bulk.
The field equation is given by
\begin{eqnarray}
\left(\partial_y^2 -3 \coth y \partial_y + 
\left( p'^2+ \frac{9}{4} \right)
- \frac{m^2 l^2}{\sinh^2 y}
\right) f_{p'}(y)= 0, \nonumber\\
\Box_4 Y_{p'}(x)= \left(p'^2+ \frac{9}{4} \right)Y_{p'}(x). 
\label{4-48}
\end{eqnarray}
Making the analytic continuation to Euclidean spacetime
and putting
\begin{equation}
p'=i \left(p+\frac{3}{2} \right),
\label{4-49}
\end{equation}
we obtain the solutions of the field equation (\ref{4-48})
\begin{eqnarray}
f_p(y)=(\sinh y)^{2} Q^{\Delta-2}_{p+1}(\cosh y), \nonumber\\
\Box_4 Y_{p}(x)=-p(p+3)Y_p(x),
\label{4-50}
\end{eqnarray}
where $Q$ is the associated Legendre function of the
second kind, $\Delta=2+\sqrt{4+m^2 l^2}$, $Y_p(x)$
is the scalar harmonics on $S^4$.
From Euclidean boundary condition, 
we have chosen the solution $f_p(y)$ which is regular 
at $y \to \infty$. Using these mode functions, 
we can write the solution of the
field equation with boundary condition $\phi(y=y_0)=\phi_B$ as
\begin{equation}
\phi(y,x)=\sum_p \frac{f_p(y)}{f_p(y_0)} Y_{p}(x) \phi_B(p).
\label{4-51}
\end{equation}
Then the generating function can be obtained by inserting
the solution ($\ref{4-51}$) into ($\ref{2-7}$) and using
the orthonormal relation of the harmonics,
\begin{equation}
\int d^4 x \sqrt{h} Y_p(x)Y_{p'}^*(x)=\delta_{p p'}.
\end{equation}
We obtain
\begin{eqnarray}
I[\phi_B] &=& \sum_p \:  \lim_{\epsilon \to 0}
[\epsilon^{-3} K_p(\epsilon)] \phi_B(p) \phi_B(-p), \nonumber\\
K_p[\epsilon] &=&  - \frac{1}{2} \frac{f'(y_0)}{f(y_0)},
\label{4-52}
\end{eqnarray}
where the field is rescaled as $\phi_B \to l^{-3/2} \phi_B$.
There is a problem in taking the limit $\epsilon \to 0$
because $K_p(\epsilon) $ diverges as $\epsilon \to 0$.
We should introduce the counter terms to regulate the
result. Since the structure of the divergence is different
for massless and massive field, we will treat them separately.

First we consider the massless case.
$K_p(\epsilon)$ can be expanded in terms of
$\epsilon$ as
\begin{eqnarray}
\epsilon^{-3} K_p(\epsilon)
&=&  \frac{1}{4}p(p+3) \epsilon^{-2} + \frac{1}{8}p(p+1)(p+2)(p+3)
\log \epsilon \nonumber\\
&+& \frac{1}{8} \left(-p(p+3)+p(p+1)(p+2)(p+3)
(\psi(p+2)+\gamma-\log 2)\right)
+ \cdot \cdot \cdot,
\label{4-54}
\end{eqnarray}
where $\psi$ is the poli-gamma function and $\gamma$
is the Euler number.
The first two terms diverge as $\epsilon \to 0$ and 
these terms should be canceled by the counter terms.
The covariant forms of the counter terms are calculated
in the Appendix using the Hamilton-Jacobi equation.
The counter terms are given by
\begin{equation}
e^{S_{CT}}=e^{S^{(2)}+S^{(4)}} ,
\label{4-55}
\end{equation}
\begin{eqnarray}
S^{(2)} &=& -\frac{1}{4} \epsilon^{-2}
\int d^4 x \sqrt{h} h^{\mu \nu}
\partial_{\mu} \phi \partial_{\nu} \phi ,\nonumber\\
S^{(4)} &=& -\frac{1}{8} \log \epsilon
\int d^4 x \sqrt{h} \left(
\frac{2}{3} R  \partial_{\mu} \phi \partial^{\mu} \phi
-2 R_{\mu \nu} \partial_{\mu} \phi \partial_{\nu} \phi
+(\Box_4 \phi)^2
\right),
\label{4-56}
\end{eqnarray}
where we have made analytic continuation of the result 
obtained in the appendix (\ref{AA-1}) to Euclidean spacetime.
At finite order of $\epsilon$,
terms  which are analytic in $p$ can be canceled by a
local counter term. We will consider
the term which cannot be canceled by a local counter term.
The correpondance in Euclidean spcaetime is given by
\begin{equation}
e^{-I[\phi_B]}=\left \langle \exp \left(\int d^4 x \sqrt{h}
{\cal O} \phi_B \right) \right \rangle.
\end{equation}
Then, the two point function of CFT can be obtained by
varying $I$ twice with respect to $\phi_B$ as
\begin{equation}
\langle {\cal O}(p) {\cal O}(-p) \rangle
= -\frac{1}{4}  p(p+1)(p+2)(p+3) \psi(p+2).
\label{4-58}
\end{equation}

Next consider the massive field. $K_p$ can be expanded
as
\begin{eqnarray}
\epsilon^{-3} K_p= \frac{1}{2}\epsilon^{-4} (\Delta -4)+\frac{1}{4} 
\epsilon^{-2}
\left( \frac{p(p+3)}{\Delta-3} - \frac{2(\Delta-4)}{\Delta-3} \right)
\nonumber\\
+ \frac{1}{2} \epsilon^{2(4-\Delta)} 
2^{-2\Delta+5} \frac{\Gamma(3- \Delta)
\Gamma(p+ \Delta)}{\Gamma(\Delta-2) \Gamma(p-\Delta+4)}
\cdot \cdot \cdot.
\label{4-59}
\end{eqnarray}
Again, the first two terms diverge as $\epsilon\to 0$.
These divergences can be canceled by the counter terms
\begin{eqnarray}
S^{(0)}&=& \frac{1}{2} \epsilon^{-4}\int d^4 x
\sqrt{h} (4-\Delta) \phi^2, \nonumber\\
S^{(2)}&=& -\frac{1}{4(\Delta-3)} \epsilon^{-2}
\int d^4 x \sqrt{h}
\left( \frac{4-\Delta}{6} R \phi^2+h^{\mu \nu}
 \partial_{\mu} \phi \partial_{\nu} \phi \right),
\label{4-60}
\end{eqnarray}
where we have made analytic continuation of the result 
in the appendix (\ref{AA-2}) to Euclidean spacetime.
There is a finite term of the order $\epsilon^{0}$. Because 
it is analytic in $p$ ($\propto (p+4-\Delta)(p-1+\Delta)
(p+1)(p+2)$), it can be canceled by a local counter term.
The leading term which is non-analytic in $p$ is the last term in
(\ref{4-59}).
Rescaling the source function $\phi_B \to \epsilon^{4-\Delta}
\phi_B $, the two point function of the CFT becomes
\begin{equation}
\langle {\cal O}(p) {\cal O}(-p) \rangle
=-2^{-2\Delta+5} \frac{\Gamma(3- \Delta)
\Gamma(p+ \Delta)}{\Gamma(\Delta-2) \Gamma(p-\Delta+4)}.
\label{4-61}
\end{equation}

The correlation function in the real spacetime is obtained
by mode summation of harmonics of $S^4$
\cite{R}
\begin{equation}
\sum_{lmn} Y_{plmn}(x)
Y^*_{p lmn}(x') =\frac{1}{4 \pi^{5/2}}
(2p+3) \Gamma \left(\frac{3}{2} \right) C^{3/2}_p (\cos \gamma_4),
\end{equation}
and the formula \cite{SUM}
\begin{equation}
\sum_p \frac{\Gamma(p+ \Delta)}{\Gamma(p-\Delta+4)}
\left(p+ \frac{3}{2} \right) \Gamma \left(\frac{3}{2} \right)
C^{3/2}_p (\cos \gamma_4)
= \frac{\Gamma(\Delta)}{\Gamma(2-\Delta)} 2^{\Delta-3} \sqrt{\pi}
(1-\cos \gamma_4)^{-\Delta},
\end{equation}
where $C$ is the Gegenbauer polynomials and $\gamma_4$ is the
angle between $x$ and $x'$.
Then the correlation function of the CFT in the real space
can be written as
\begin{eqnarray}
\langle {\cal O} (x) {\cal O} (x') \rangle
&=& \sum_p \langle {\cal O}(p) {\cal O}(-p) \rangle Y_p(x) 
Y^*_p(x')
\nonumber\\
&=& \frac{2}{\pi^2}
\frac{(\Delta-2) \Gamma(\Delta)}{\Gamma(\Delta-2)}
\frac{1}{\sigma_E^{2 \Delta}},
\label{4-64}
\end{eqnarray}
where $\sigma_E^2=2(1-\cos \gamma_4)$ 
is the invariant length of $S^4$.
By making analytic continuation to Lorentzian spacetime,
we find the result agrees with (\ref{4-46}).

Now we examine the nature of the CFT in de Sitter spacetime.
Making analytic continuation to the Lorentzian spacetime,
the closed de Sitter spacetime
\begin{equation}
d s_4^2=\frac{1}{H^2}(-dt^2 +\cosh^2 H t \: d \Omega_3^2),
\end{equation}
is obtained and the invariant length becomes
\begin{equation}
\sigma_E^2 \to \frac{2}{H^2} \left
(1+ \sinh Ht \sinh H t'-\cosh H t \cosh H t' \cos \gamma_3
\right),
\end{equation}
where $\gamma_3$ is the angle between $(\Omega_3,\Omega_3')$.
For equal spatial point at which $\cos \gamma_3=1$, the two-point 
function is given by
\begin{equation}
\langle {\cal O} (t) {\cal O} (0) \rangle  \propto
\left(\frac{H}{\sinh \left(\frac{H t}{2} \right)} \right)^{2 \Delta}.
\end{equation}
We find this two-point function has periodicity in imaginary
time $it \to it + 2 \pi/H$. Thus we can interpret it as
the thermal Green function with temperature $T=H/2 \pi$.

%%%%%%%%%%%%%%%%%%%%%%%%%%%%%%%%%%%%%%%%%%%%%%%%%%%%%%%%%%
%%%%%%%%%%%%%%%%%%%%%%%%%%%%%%%%%%%%%%%%%%%%%%%%%%%%%%%%%%

\section{CFT in FRW universe}
We extend the calculation to FRW universe.
Let us consider a coordinate transformation 
\begin{equation}
z=f(u)-g(v), \quad \tau=f(u)+g(v),
\label{2-18}
\end{equation}
with arbitrary functions $f(u)$ and $g(v)$ where $u=t-y$ and $v=t+y$
\cite{KS1}.
Then the metric (\ref{2-9}) becomes
\begin{equation}
l^{-2} ds^2
=e^{2 \beta(y,t)}(dy^2-dt^2)+e^{2\alpha(y,t)}(\delta_{ij} dx^i dx^j).
\label{2-19}
\end{equation}
Here $\alpha(y,t)$ and $\beta(y,t)$ are given by
\begin{eqnarray}
e^{2\beta(y,t)}=4 \frac{f'(u)g'(v)}{(f(u)-g(v))^2} ,\quad
e^{2\alpha(y,t)}=\frac{1}{(f(u)-g(v))^2}.
\label{2-20}
\end{eqnarray}
The $y=0$ slicing can be the FRW spacetime
\begin{equation}
l^{-2} ds_d^2= -e^{2 \beta_0(t)} dt^2 + e^{2 \alpha_0(t)}
\delta_{ij} dx^i dx^j,
\label{2-21}
\end{equation}
with scale factor $e^{\alpha_0}$ where $\alpha_0(t)=
\alpha(0,t)$ and $\beta_0(t)=\beta(0,t)$.

As in the de Sitter spacetime, 
some efforts to take the limit to
the boundary are needed, 
because the FRW spacetime can be realized on only one
slicing of the AdS spacetime, that is $y=0$.
We take the following procedure.
First we take the limit $y \to 0$.
Since $z(0,t)=e^{-\alpha_0(t)}$, 
further liming procedure is needed to go to
the boundary $z \to 0$. The scale factor $e^{\alpha_0}$
has one integration of the constant. Then in general,
the scale factor can be written as $e^{\alpha_0}=
a(t)/a_{\star}$. If we take $a_{\star} \to 0$,
$y=0$ slicing effectively goes to the boundary
$z(0,t)=a_{\star}/a(t) \to 0$. 
Thus we can take $\epsilon=a_{\star}$.
Introducing the conformal time $\eta$, 
the d-dimensional line element is given by
\begin{equation}
ds_4^2= (l/a_{\star})^2
a(\eta)^2 (-d \eta^2 +\delta_{ij} dx^i dx^j).
\end{equation}
Thus in order to fix the scale factor of the d-dimensional
spacetime, we should take $\l \to 0$ as $a_{\star} \to 0$. 

We use the invariant Green function in constructing the
bulk-boundary propagator. The explicit form of the coordinate
transformation is needed. 
The coordinate transformation $f(u)$ and $g(v)$
can be determined by specifying the scale factor $e^{\alpha_0(t)}$
and fixing the gauge $e^{\beta_0(t)}$.
For example, we take $d=4$ and
\begin{equation}
e^{-\alpha_0(t)} =f(t)-g(t)= a_{\star}
t^{-\frac{2}{3(1+w)}}, \quad
e^{2 \beta_0(t)} =4 \frac{f'(t)g'(t)}{(f(t)-g(t))^2}= 1,
\label{2-24}
\end{equation}
where $w$ is a constant parameter and can be regarded as 
the barotropic parameter of the matter 
which dominates the universe.
Note that the conformal time is defined by
\begin{equation}
\eta=a_{\star} t^{-\frac{2}{3(1+w)}+1} \frac{3(1+w)}{1+3w}.
\end{equation}
Thus, in order to fix the conformal time in the limiting procedure
$a_{\star} \to 0$, we should take $t \gg 1$.
The equations (\ref{2-24}) give the
simultaneous first order differential equations for
$f(t)$ and $g(t)$;
\begin{eqnarray}
f'(t) &=&\frac{a_{\star}}{3(1+w)} t^{-\frac{5+3w}{3(1+w)}}
\left( -1+ \frac{3}{2}(1+w)t \right), \nonumber\\
f(t) &=& g(t)+ a_{\star} t^{-\frac{2}{3(1+w)}},
\label{2-25}
\end{eqnarray}
where we have used $t \gg 1$.
Once the functions $f(t)$ and $g(t)$ are obtained,
the coordinate transformation is determined by
replacing $f(t) \to f(u)$ and $g(t) \to g(v)$ as
\begin{eqnarray}
f(u) &=& \frac{a_{\star}}{2} u^{-
\frac{2}{3(1+w)}}\left(
1+ \frac{3(1+w)}{1+3 w} u
 \right), \nonumber\\
g(v) &=& \frac{a_{\star}}{2} v^{-
\frac{2}{3(1+w)}} \left(
-1+ \frac{3(1+w)}{1+3 w} v
 \right).
 \label{2-26}
\end{eqnarray}
Then $z$ and $\tau$ can be written in terms of $y$ and $t$ and
we can show the following relations;
\begin{eqnarray}
\lim_{a_{\star} \to 0,y \to 0}
\tau &=& \lim_{a_{\star} \to 0,y \to 0}  (f(u)+g(v))=
a_{\star}t^{-2/3(1+w)+1} \frac{3(1+w)}{1+3w}  = \eta,
\nonumber\\
\lim_{a_{\star} \to 0,y \to 0} \frac{\partial z(y,t)}{\partial y}
&=& -  \lim_{a_{\star} \to 0,y \to 0}
(f'(u)+g'(v))=- a_{\star} t^{-2/3(1+w)}=
-\frac{a_{\star}}{a(\eta)}.
\label{2-27}
\end{eqnarray}
Using these relations, we obtain 
\begin{eqnarray}
\lim_{a_{\star} \to 0,y \to 0} 
\frac{\partial^2}{\partial y \partial y'}G(y,x;y'x') 
&=& \lim_{a_{\star} \to 0,y \to 0} 
\left(\frac{\partial z}{\partial y} 
\frac{\partial z'}{\partial y'} \right)
\frac{\partial^2}{\partial z \partial z'}G (y,x,;y',x')
\nonumber\\
&=& G_0  
2^{2 \Delta} \Delta^2 a_{\star}^{2 \Delta}
\left( \frac{a(\eta)^{-1} a(\eta')^{-1}}{-(\eta-\eta')^2+
\vert x^i -x'^i \vert^2} \right)^{\Delta},
\end{eqnarray}
where we have used the fact $\partial^2 G/\partial \tau
\partial \tau' $ is the higher order of $a_{\star}$.
Then using $\sqrt{-\gamma}=(a_{\star}/l)^{-d}\sqrt{-h}$ and
$n^A(\partial/\partial X^A)=-\partial/\partial y$, 
(\ref{2-8}) becomes
\begin{eqnarray}
I[\phi_B] &=& \frac{1}{2} \int_{{\cal B}}
d^d x d^d x' \sqrt{-h(x)} \sqrt{-h(x')}
\phi_B(x) \phi_B(x') a_{\star}^{2(\Delta-d)} l^{d-1} \nonumber\\ 
& \times &
\left(
G_0 2^{2 \Delta} \Delta^2
\frac{a(\eta)^{-\Delta} a(\eta')^{-\Delta}}
{\vert x-x' \vert^{2 \Delta}} \right),
\end{eqnarray}
where we denote $(a_{\star}/l)^2 ds^2_4= h_{\mu \nu}
dx^{\mu} dx^{\nu}$ and $ \vert x-x' \vert^{2}=-(\eta-\eta')^2
+ \vert x^i-x'^i \vert^2$.
Putting $a_{\star}=\epsilon$ and 
renormalizing $\phi_B$ as in the Minkowski case (\ref{2-12}), 
we get
\begin{eqnarray}
\langle {\cal O}(x) {\cal O}(x') \rangle_{FRW}
= C_2 \frac{a(\eta)^{-\Delta} a(\eta')^{-\Delta}}
{\vert x-x' \vert^{2 \Delta}}.
\label{2-30}
\end{eqnarray}
This two-point function is similar to the two-point function
of quantum field in conformal vacuum. The two-point function 
$D(x,x)$ of conformally invariant field with conformal 
dimension $\Delta$ in conformally flat 
spacetime can be written as
\begin{equation}
D(x,x')=\Omega^{\Delta}(x) D_{F}(x,x') \Omega^{\Delta}(x')
\label{2-31}
\end{equation}
where $D_{F}(x,x')$ is the two-point function in flat spacetime
and we denote the conformally flat spacetime as $h_{\mu \nu}
=\Omega^{-2} \eta_{\mu \nu}$.  The FRW universe is given by
$\Omega(x)=a(\eta)^{-1}$. 
Comparing (\ref{2-13}) and (\ref{2-30}), we see the relation
(\ref{2-31}) actually holds with $\Delta$ as the conformal 
dimension.

As in the de Sitter spacetime, the CFT in various states
can be obtained from bulk-boundary propagators with
appropriate boundary conditions. The solution of the field 
equation in the bulk is written as
\begin{equation}
\phi(y,t,x^i)= \int d p \: G_{{\cal B}}(y,t,x^i;p) \phi_B(p).
\end{equation}
The bulk-boundary propagator $G_{{\cal B}}$ satisfies 
the field equation and
\begin{equation}
\lim_{y \to 0, a_{\star} = \epsilon} G_{{\cal B}}(y,t,x^i;p)
=Y_{p}(t,x^i),
\end{equation}
where $Y_p$ is the harmonics in FRW universe. Appropriate choice
of the harmonics will lead to the CFT in desired vacuum state.

%%%%%%%%%%%%%%%%%%%%%%%%%%%%%%%%%%%%%%%%%%%%%%%%%%%%%%%%%%%%%%%%%
%%%%%%%%%%%%%%%%%%%%%%%%%%%%%%%%%%%%%%%%%%%%%%%%%%%%%%%%%%%%%%%%%
%%%%%%%%%%%%%%%%%%%%%%%%%%%%%%%%%%%%%%%%%%%%%%%%%%%%%%%%%%%%%%%%%
\section{Discussions}
We have developed a formalism to calculate the effective 
action of the strongly coupled CFT
in curved spacetime from AdS/CFT correspondence.
Using the fact that the de Sitter spacetime and FRW universe 
can be embedded into the AdS spacetime, the effective action 
of scalar fields was derived in those spacetime.
Recently Hawking et. al. used
the effective action derived via AdS/CFT correspondence 
and calculated  
the correction of tensor propagators by conformal matter 
during trace anomaly driven inflation \cite{H2}.
Their calculations are limited to exactly de Sitter spacetime.
Our formalism may be useful to extend their analysis to
more realistic situations. 

Further extensions of the formalism can be considered.
First, AdS spacetime itself can be embedded in the
higher dimensional AdS spacetime \cite{ad}. The coordinate
transformation from Poincare coordinate
\begin{equation} 
z = \eta \cos y , \quad x_1 = \eta \sin y ,
\end{equation}
gives  
\begin{equation}
ds^2= \left(\frac{l^2}{\cos^2 y}\right) 
\left(d y^2+ \frac{1}{\eta^2} (d \eta^2
-d \tau^2 + \delta_{ij} d x^i d x^j) \right).
\end{equation}
At $\cos y_0= L^{-1} l$, the $AdS_d$ with 
the curvature radius $L$ is embedded.
Thus, the CFT in AdS spacetime can be learned from
the classical theory in the AdS bulk. 
So far, we investigate the pure AdS spacetime
as the geometry of the
bulk. However, more general spacetime with negative cosmological
constant can be considered. It is known that
\begin{equation}
ds^2= \left(\frac{l^2}{y^2} \right)(dy^2+
g_{\mu \nu} dx^{\mu} dx^{\nu}),
\end{equation}
with any Ricci flat metric $g_{\mu \nu}$
and
\begin{equation}
ds^2= \left(\frac{l^2}{\sinh^2 y} \right)(dy^2+
\sigma_{\mu \nu} dx^{\mu} dx^{\nu}),
\end{equation}
with any metric $\sigma_{\mu \nu}$ with positive cosmological
constant are solutions of $d+1$ dimensional 
spacetime with negative cosmological constant.
Thus, it seems to be possible to extend the conjecture that
classical theory in these bulk spacetime is dual to quantum
field theory on its boundary. The interesting point is that
the calculation done in section 3.2 can be straightforwardly
extended because $y$ dependence of the mode function is the
same with the one in pure AdS spacetime. Because these spacetime
are in general asymptotically non AdS spacetime, it
would be challenging to identify the dual field theory.

\acknowledgments{
We would like to thank S. Kobayashi for helpful discussions.
We would like to thank Feng -Li Lin for useful comments
and for sending us their unpublished notes on the related issue. 
The work of K.K. was supported by JSPS Research Fellowships for
Young Scientist No.4687. 
}

\appendix
\section{Derivation of the counter term}
In this appendix, we present the derivation of the
covariant form of the
counter terms via Hamilton-Jacobi equation.
Other derivations of the counter terms can be seen 
in \cite{C-G} for pure gravity and in \cite{C-S} 
for scalar field theory.

\subsection{Hamilton-Jacobi equation}
To derive the counter terms, we should know 
the behaviour of the classical action $I$
at the boundary. 
One of the easiest way to obtain $I$ is using the
Hamilton-Jacobi (H-J) equation \cite{HJ}.
Let us consider the $(d+1)$-dimensional action
\begin{equation}
S =\int d^{d+1} x \sqrt{-g} \left(
\frac{1}{2 \kappa} \left( \mbox{}^{(d+1)}R- 2 \Lambda
\right)-
\frac{1}{2} \partial_{\mu} \phi \partial^{\mu} \phi
-\frac{1}{2} m^2 \phi^2 \right),
\end{equation}
where $ \mbox{}^{(d+1)}R$ is the Ricci tensor in 
$d+1$-dimensional spacetime.
The line element is denoted as
\begin{equation}
ds^2=(N^2 + \gamma_{\mu \nu} N^{\mu} N^{\nu}) du^2 +
2 N_{\mu} du dx^{\mu} + \gamma_{\mu \nu} dx^{\mu} dx^{\nu},
\end{equation}
where $N$ is the lapse function, $N^{\mu}$ is the shift
function and $\gamma_{\mu \nu}$ is the d-dimensional
metric. By defining the conjugate momentum
\begin{equation}
\pi^{\mu \nu } = \frac{\delta S}{\delta \dot{\gamma}_{\mu \nu}}, \quad
\pi_{\phi} = \frac{\delta S}{\delta \dot{\phi}},
\end{equation}
the Hamiltonian form of the action
\begin{equation}
S= \int d^{d+1}x  \left( \pi^{\mu \nu} \dot{\gamma}_{\mu \nu}
+ \pi_{\phi} \dot{\phi} - N {\cal H}- N^{\mu} {\cal H}_{\mu} \right),
\end{equation}
is obtained where
\begin{eqnarray}
{\cal H} &=& -\frac{2 \kappa}{\sqrt{-\gamma}} \pi^{\mu \nu}
\pi^{\lambda \rho} 
\left( \gamma_{\mu \rho } \gamma_{\nu \lambda}-
\frac{1}{d-1} \gamma_{\mu \nu } \gamma_{\lambda \rho} \right)
-\frac{\sqrt{-\gamma}}{2 \kappa} (R-2 \Lambda) \nonumber\\
&-& \frac{1}{2 \sqrt{-\gamma}} \pi_{\phi}^2+ \frac{1}{2}
\sqrt{-\gamma} \gamma^{\mu \nu} \partial_{\mu} \phi
\partial_{\nu} \phi +\frac{1}{2 \sqrt{-\gamma}} m^2 \phi^2,
\nonumber\\
{\cal H}_{\mu} &=& -2 \pi_{\mu \:\:\:; \nu}^{\nu}+
\pi_{\phi} \phi_{,\mu},
\end{eqnarray}
$R$ is the Ricci tensor of $\gamma_{\mu \nu}$ 
and $;$ denotes the covariant derivative with respect to 
$\gamma_{\mu \nu}$.
Variation of the action with respect to the momentum gives
the equation of the motion for $\gamma_{\mu \nu}$ and $\phi$;
\begin{eqnarray}
\dot{\gamma}_{\mu \nu} -N_{\mu; \nu}-N_{\nu; \mu}
&=& -N \frac{4 \kappa}{\sqrt{-\gamma}} \pi^{\lambda \rho}
\left(\gamma_{\mu \rho}\gamma_{\nu \lambda}-
\frac{1}{d-1} \gamma_{\mu \nu} \gamma_{\lambda \rho} \right),
\nonumber\\
\dot{\phi}-N^{\mu} \phi_{, \mu} 
&=& -\frac{N}{\sqrt{-\gamma}} \pi_{\phi},
\label{A-0}
\end{eqnarray}
where $\cdot$ denotes the derivative with respect
to $u$.
The H-J equation is obtained by putting
\begin{equation}
\pi^{\mu \nu }= \frac{\delta I}{\delta \gamma_{\mu \nu}} ,
\quad \pi_{\phi}=\frac{\delta I}{\delta \phi},
\label{A-0-1}
\end{equation}
where $I$ is the classical action of the system
which is obtained by inserting the solutions of 
field equations into $S$. The H-J equation is
given by
\begin{eqnarray}
{\cal H} \left(\frac{\delta I}{\delta \gamma_{\mu \nu}},
\frac{\delta I}{\delta \phi},\gamma_{\mu \nu},\phi
\right)
&=& -\frac{2 \kappa}{\sqrt{-\gamma}}
\frac{\delta I}{\delta \gamma_{\mu \nu}}
\frac{\delta I}{\delta \gamma_{\lambda \rho}}
\left( \gamma_{\mu \rho } \gamma_{\nu \lambda}-
\frac{1}{d-1} \gamma_{\mu \nu } \gamma_{\lambda \rho} \right)
-\frac{\sqrt{-\gamma}}{2 \kappa}(R -2 \Lambda) \nonumber\\
&-& \frac{1}{2 \sqrt{-\gamma}} \left(\frac{\delta I}
{\delta \phi} \right)^2
+\frac{1}{2} \sqrt{-\gamma} \gamma^{\mu \nu}
\partial_{\mu} \phi \partial_{\nu} \phi +\sqrt{-\gamma}
\frac{1}{2} m^2 \phi^2=0.
\end{eqnarray}
In the following calculations, we use the gauge
$N=1$ and $N^{\mu}=0$. However, because the H-J equation 
does not contain neither Lapse function $N$ nor shift 
function $N_{\mu}$, 
the solution of the H-J equation does not depend
on the choice of the slicing in ($d+1$) spacetime.

From (\ref{4-56}) and (\ref{4-60}),
we notice that the expansion with respect to $\epsilon$ is also
expansion with respect to the number of the derivatives
of d-dimensional spacetime. Thus we will obtain $I$
order by order with respect to the number of the derivatives
of d-dimensional spacetime,
\begin{equation}
I=I^{(0)}+I^{(2)} +\cdot \cdot \cdot .
\end{equation}
We will use fixed background approximation,
that is, we assume the motion of the scalar field
does not affects the geometry of the spacetime.
We will treat separately a massless ($m=0$) field 
and a massive $m > 0$ field.

\subsection{Solution of H-J equation -massless case}
The Hamiltonian constraint can be also
expanded in terms of the number of the
derivatives as
\begin{equation}
{\cal H}={\cal H}^{(0)} +
{\cal H}^{(2)} + \cdot \cdot \cdot.
\end{equation}
\hspace{1cm}\\
(1) 0-th order\\
At the 0-th order, the H-J equation is given by
\begin{equation}
{\cal H}^{(0)} =
 -\frac{2 \kappa}{\sqrt{-\gamma}}
\frac{\delta I^{(0)}}{\delta \gamma_{\mu \nu}}
\frac{\delta I^{(0)}}{\delta \gamma_{\lambda \rho}}
\left( \gamma_{\mu \rho} \gamma_{\nu \lambda}-
\frac{1}{d-1} \gamma_{\mu \nu } \gamma_{\lambda \rho} \right)
+ \frac{\sqrt{-\gamma}}{\kappa} \Lambda=0,
\label{A-1}
\end{equation}
where we take $\delta I^{(0)}/\delta \phi =0$ from
fixed background approximation.
Then putting the solution of the form
\begin{equation}
I^{(0)}= 2 A \int d^d x \sqrt{-\gamma},
\end{equation}
we have
\begin{equation}
A=A_{\pm}= \pm \frac{d-1}{2 \kappa l},
\label{A-1-1}
\end{equation}
where we have used $\Lambda=-d(d-1)/2 l^2$.
We will take the solution $A=A_{+}$. 
Inserting the solution into (\ref{A-0-1}),
we can obtain the solutions of the momentum.
Then the equation of the motion (\ref{A-0}) becomes
\begin{equation}
\dot{\gamma}_{\mu \nu}= \frac{2}{l}  \gamma_{\mu \nu}.
\end{equation}
Thus the solution is obtained as
\begin{eqnarray}
\gamma_{\mu \nu} &=& \Omega^2(u) f_{\mu \nu},
\quad \Omega(u)=e^{u/l}.
\label{A-2}
\end{eqnarray}
where $f_{\mu \nu}$ is independent of $u$.
\hspace{1cm}\\
(2) 2-nd order \\
The H-J equation is given by
\begin{equation}
{\cal H}^{(2)} = \frac{2}{l}
\gamma_{\mu \nu}
\frac{\delta I^{(2)}}{\delta \gamma_{\mu \nu}}
-\frac{\sqrt{-\gamma}}{2 \kappa}
R + \frac{1}{2} \sqrt{-\gamma} \gamma^{\mu \nu}
\partial_{\nu} \phi \partial_{\mu} \phi=0.
\label{A-2-2}
\end{equation}
Using the solution ($\ref{A-2}$), we can write
\begin{equation}
\frac{\delta I^{(2)}}{\delta u}= 
\frac{2}{l} \gamma_{\mu \nu} \frac{\delta I^{(2)}}
{\delta \gamma_{\mu \nu}}.
\end{equation}
By the standard result of the conformal transformation
\begin{equation}
R(\gamma_{\mu \nu})=\Omega^{-2} R(f_{\mu \nu}), \quad
\sqrt{-\gamma}= \Omega^d \sqrt{-f},
\end{equation}
the H-J equation becomes
\begin{eqnarray}
\frac{\delta I^{(2)}}{\delta u}
&=&  \frac{1}{2 \kappa} \sqrt{-\gamma}
\left(
 R(\gamma_{\mu \nu})
- \kappa \gamma^{\mu \nu} \partial_{\mu} \phi
\partial_{\nu} \phi \right) \nonumber\\
&=& \frac{1}{2 \kappa} \sqrt{-f}
\left( R(f_{\mu \nu})- \kappa f^{\mu \nu}
\partial_{\mu} \phi \partial_{\nu} \phi \right)
\Omega(u)^{d-2}.
\end{eqnarray}
Apart from $\Omega(u)^{d-2}$, 
the right hand side of the equation is
independent of $u$. Thus we can immediately
integrate the equation. The solution is given by
\begin{equation}
I^{(2)}=\frac{l}{d-2}
\int d^d x \sqrt{-\gamma} \left( \frac{1}{2 \kappa} R
-\frac{1}{2} \gamma^{\mu \nu} \partial_{\mu} \phi
\partial_{\nu} \phi \right).
\label{A-3}
\end{equation}
\hspace{1cm}\\
(2) 4-th order\\
The H-J equation is given by
\begin{eqnarray}
{\cal H}^{(4)} &=& \frac{2}{l} 
\gamma_{\mu \nu}
\frac{\delta I^{(4)}}{\delta \gamma_{\mu \nu}}
- \frac{2\kappa}{\sqrt{-\gamma}}
\frac{\delta I^{(2)}}{\delta \gamma_{\mu \nu}}
\frac{\delta I^{(2)}}{\delta \gamma_{\lambda \rho}}
\left(
\gamma_{\mu \rho} \gamma_{\nu \lambda} -
\frac{1}{d-1} \gamma_{\mu \nu} \gamma_{\rho \lambda}
\right) \nonumber\\
&-& \frac{1}{2 \sqrt{-\gamma}} \left(
\frac{\delta I^{(2)}}{\delta \phi} \right)^2=0.
\end{eqnarray}
Inserting the solution (\ref{A-2}) and $I^{(2)}$
(\ref{A-3}), the H-J equation becomes 
\begin{eqnarray}
\frac{\delta I^{(4)}}{ \delta u}
&=& \Omega(u)^{d-4} F_4[f,\phi] ,\nonumber\\
F_4[f,\phi] &=&
\frac{l^2}{2 \kappa (d-2)^2} \sqrt{-f}
\left(R_{\mu \nu} R^{\mu \nu}-\frac{d}{4(d-1)}
R^2 \right. \nonumber\\
&+&  \left.
 \frac{d}{2(d-1)} \kappa R
\partial_{\mu} \phi \partial^{\mu} \phi
-2 \kappa R^{\mu \nu} \partial_{\mu} \phi
\partial_{\nu} \phi + \kappa(\Box \phi)^2 
+ \frac{3d-4}{4(d-1)}
\kappa^2 (\partial_{\mu} \phi \partial^{\mu} \phi)^2
\right). \nonumber\\
\end{eqnarray}
Again, we observe that right-hand side of the equation
is independent of $u$ apart from $\Omega(u)^{d-4}$.
Then we can easily integrate the equation to obtain
\begin{equation}
I^{(4)}=
\frac{l^3}{2 \kappa (d-2)^2 (d-4)}
\int d^d x \sqrt{-\gamma}
F_4[\gamma,\phi].
\end{equation}
\hspace{1cm} \\
(3)higher order \\
The higher order solution ${\cal H}^{(2n)}$ ($n \geq 2$) 
can be obtained recursively.
The H-J equation can be written as
\begin{equation}
{\cal H}^{(2n)} =
\frac{\delta I^{(2n)}}
{\delta u}
-F_{2n}[\gamma,\phi]=0 ,
\end{equation}
\begin{eqnarray}
F_{2n}[\gamma,\phi] &=&
\frac{2 \kappa}{\sqrt{-\gamma}} \sum_{p=1}^{n-1}
\frac{\delta I^{(2p)}}{\delta \gamma_{\mu \nu}}
\frac{\delta I^{(2n-2p)}}{\delta \gamma_{\lambda \rho}}
\left(
\gamma_{\mu \rho} \gamma_{\nu \lambda} -
\frac{1}{d-1} \gamma_{\mu \nu} \gamma_{\rho \lambda}
\right) \nonumber\\
&+& \frac{1}{2 \sqrt{-\gamma}} \sum_{p=1}^{n-1}
\frac{\delta I^{(2p)}}{\delta \phi}
\frac{\delta I^{(2n-2p)}}{\delta \phi}. 
\end{eqnarray}
Using the solution ($\ref{A-2}$), we find 
$F_{2n}[\gamma,\phi]=\Omega^{d-2n}(u)
F_{2n}[f,\phi]$. Then integration of the
equation yields
\begin{equation}
I^{(2n)}
= \frac{l}{d-2n} F_{2n}[\gamma,\phi].
\label{A-3-1}
\end{equation}

\subsection{Solution of H-J equation -massive case}
(1) 0-th order\\
We take the following ansatz for 0-th order solution
\begin{equation}
I^{(0)}= 2 \int d^d x \sqrt{-\gamma} H(\phi).
\end{equation}
The H-J equation is gives by
\begin{equation}
2 \kappa \frac{d}{d-1} H^2(\phi)
-2 \left(
\frac{\partial H(\phi)}{\partial \phi} 
\right)^2 +\frac{\Lambda}{\kappa}
+\frac{1}{2} m^2 \phi^2=0.
\end{equation}
From the background approximation, we
assume 
\begin{equation}
H(\phi)=A+ \frac{1}{2} B \phi^2,
\end{equation}
where $A \gg B \phi^2 $ and $A$ is given by
(\ref{A-1-1}).
Then we obtain
\begin{equation}
B=\frac{d \pm \sqrt{d^2 +4 m^2 l^2}}{4 l} \equiv
\frac{d- \Delta_{\mp}}{2l},
\end{equation}
We will pick the solution $\Delta_{+}=\Delta$.
The equation of motion ($\ref{A-0}$) becomes
\begin{eqnarray}
\dot{\gamma} &=& \frac{2}{l} \gamma_{\mu \nu},
\nonumber\\
\dot{\phi} &=& -\frac{d-\Delta}{l} \phi.
\end{eqnarray}
The solution can be obtained as
\begin{eqnarray}
\gamma_{\mu \nu} &=& \Omega^2(u) f_{\mu \nu},
\quad \Omega(u)=e^{u/l}, \nonumber\\
\phi &=& \Pi(u) C ,\quad
\Pi(u)=e^{-(d-\Delta)u/l}.
\label{A-5}
\end{eqnarray}
where $C$ is independent of $u$.
\hspace{1cm}\\
In the following, we divide the classical action as
\begin{equation}
I=I_0(\gamma_{\mu \nu})+I_1(\gamma_{\mu \nu},\phi),
\end{equation}
and assume $I_0 \gg I_1$. \\
(2)2-nd order\\
Inserting the solution of the first order, 
the H-J equation becomes
\begin{equation}
\frac{2}{l}  \gamma_{\mu \nu}
\frac{\delta I_0^{(2)}}{\delta \gamma_{\mu \nu}}
-\sqrt{-\gamma} \frac{R}{2 \kappa}=0,
\end{equation}
\begin{equation}
\frac{2}{l}  \gamma_{\mu \nu}
\frac{\delta I_1^{(2)}}{\delta \gamma_{\mu \nu}}
-\frac{d-\Delta}{l} \phi \frac{\delta I_1^{(2)}}{\delta \phi}
= -\frac{1}{2} \sqrt{-\gamma} \gamma^{\mu \nu}
\partial_{\mu} \phi \partial_{\nu} \phi
-\frac{d-\Delta}{4(d-1)} \sqrt{-\gamma}  R \phi^2.
\end{equation}
The equation for $I_0^{(2)}$ is the same with
(\ref{A-2-2}) without $\phi$ dependent term.
Then $I_0^{(2)}$ is given by 
\begin{equation}
I_0^{(2)}=\frac{l}{2 \kappa(d-2)} \int d^d x \sqrt{-\gamma} R.
\end{equation}
We can use the solution ($\ref{A-5}$) to write
\begin{equation}
\frac{\delta I_1^{(2)}}{\delta u}=
\frac{2}{l}  \gamma_{\mu \nu}
\frac{\delta I_1^{(2)}}{\delta \gamma_{\mu \nu}}
-\frac{d-\Delta}{l} \phi \frac{\delta I_1^{(2)}}{\delta \phi}.
\end{equation}
Then the H-J equation gives
\begin{equation}
\frac{\delta I_1^{(2)}}{\delta u}=-
\left[ \frac{d-\Delta}{4(d-1)} \sqrt{-f} R(f) C^2
+\frac{1}{2} \sqrt{-f} f^{\mu \nu} \partial_{\mu} C
\partial_{\nu} C \right] \Omega(u)^{d-2} \Pi(u)^2.
\end{equation}
Again apart from the term $\Omega(u)^{d-2}
\Pi(u)^2$, the right-hand
side is independent of $u$. Then integration can be
done to give
\begin{equation}
I_1^{(2)}=-\frac{l}{2 \Delta-d-2} \int
d^d x \sqrt{-\gamma} \left(
\frac{d-\Delta}{4(d-1)} R \phi^2
+ \frac{1}{2} \gamma^{\mu \nu} \partial_{\mu}
\phi \partial_{\nu} \phi \right).
\end{equation}
\hspace{1cm}\\
(3) higher order \\
The H-J equation for
$I^{(2n)}_0$ is the same with (\ref{A-3-1}) without
$\phi$ dependent terms
\begin{equation}
I^{(2n)}_0= 
\frac{l}{d-2n} F_{2n}[\gamma].
\end{equation}
Then the solution can be found easily. 
By using the solution (\ref{A-5}), the
H-J equation for $I^{(2n)}_1$ is written as
\begin{equation}
{\cal H}^{(2n)}_1=
\frac{\delta I_1^{(2n)}}{\delta u}
-G_{2n}[\gamma,\phi]=0,
\end{equation}
\begin{eqnarray}
G_{2n}[\gamma,\phi]
&=& \frac{4 \kappa}{\sqrt{-\gamma}} \sum_{p=1}^{n}
\frac{\delta I^{(2p)}_0}{\delta \gamma_{\mu \nu}}
\frac{\delta I^{(2n-2p)}_1}{\delta \gamma_{\lambda \rho}}
\left(
\gamma_{\mu \rho} \gamma_{\nu \lambda} -
\frac{1}{d-1} \gamma_{\mu \nu} \gamma_{\rho \lambda}
\right) \nonumber\\
&+& \frac{1}{2 \sqrt{-\gamma}} \sum_{p=1}^{n-1}
\frac{\delta I^{(2p)}_1}{\delta \phi}
\frac{\delta I^{(2n-2p)}_1}{\delta \phi} .
\end{eqnarray}
Using the solution (\ref{A-5}), we find
$G_{2n}[\gamma,\phi] = \Omega(u)^{d-2n} \Pi(u)^2
G_{2n}[f,C]$.
Then integration of the equation gives  
\begin{equation}
I^{(2n)}
=\frac{l}{2 \Delta-d-2n} G_{2n}[\gamma,\phi].
\end{equation}

\subsection{counter term}
The counter terms are defined by
\begin{equation}
S^{(0)}+S^{(2)}+ \cdot \cdot \cdot =-(
I^{(0)}+ I^{(2)} + \cdot \cdot \cdot).
\end{equation}
We summarise the result up to terms
which are necessary for the calculation with $d=4$;
\hspace{1cm}\\
(1)massless case
\begin{eqnarray}
S^{(0)}&=& -\frac{d-1}{\kappa l}
\int d^d x \sqrt{-\gamma} ,\nonumber\\
S^{(2)}&=& -\frac{l}{d-2}
\int d^d x \sqrt{-\gamma} \left( \frac{1}{2 \kappa} R
-\frac{1}{2} \gamma^{\mu \nu} \partial_{\mu} \phi
\partial_{\nu} \phi \right), \nonumber\\
S^{(4)}&=&
-\frac{l^3}{2 (d-2)^2 (d-4)} \int d^d x
\sqrt{-\gamma}
\left( \frac{1}{\kappa} \left(
R_{\mu \nu} R^{\mu \nu}-\frac{d}{4(d-1)}
R^2 \right)\right. \nonumber\\
&+&  \left.
 \frac{d}{2(d-1)} R
\partial_{\mu} \phi \partial^{\mu} \phi
-2  R^{\mu \nu} \partial_{\mu} \phi
\partial_{\nu} \phi + (
\Box \phi)^2 + \frac{3d-4}{4(d-1)}
\kappa (\partial_{\mu} \phi \partial^{\mu} \phi)^2
\right). \nonumber\\
\label{AA-1}
\end{eqnarray}
(1)massive case
\begin{eqnarray}
S^{(0)} &=& -\frac{d-1}{\kappa l}
\int d^d x \sqrt{-\gamma} -
\frac{d-\Delta}{2 l} \int d^d x \sqrt{-\gamma} \phi^2,
\nonumber\\
\!\!\!\!\!\!
S^{(2)}&=& -\frac{l}{d-2}  \int d^d x \sqrt{-\gamma}
\frac{1}{2 \kappa} R \nonumber\\
&+& \frac{l}{2 \Delta-d-2} \int
d^d x \sqrt{-\gamma} \left(
\frac{d-\Delta}{4(d-1)} R \phi^2
+ \frac{1}{2} \gamma^{\mu \nu} \partial_{\mu}
\phi \partial_{\nu} \phi \right). \nonumber\\
\label{AA-2}
\end{eqnarray}

Let us take $d=4$. 
For massless case, $S^{(4)}$ diverges at $d=4$.
We replace $1/(d-4)$ by $-\log \epsilon$.
The counter terms can be rewritten using the
metric of $d$-dimensional spacetime $h_{\mu \nu}$ 
which is defined as (see (\ref{4-46-0})),
\begin{equation}
\gamma_{\mu \nu}= \left(\frac{l}{\epsilon}\right)^2
h_{\mu \nu}.
\end{equation}
Rescaling the
field as $\phi \to l^{-3/2} \phi$, we obtain
\hspace{1cm}\\
(1)massless case
\begin{eqnarray}
S^{(2)} &=& \frac{1}{4} \epsilon^{-2}
\int d^4 x \sqrt{-h} h^{\mu \nu}
\partial_{\mu} \phi \partial_{\nu} \phi ,\nonumber\\
S^{(4)} &=& \frac{1}{8} \log \epsilon
\int d^4 x \sqrt{-h} \left(
\frac{2}{3} R  \partial_{\mu} \phi \partial^{\mu} \phi
-2 R^{\mu \nu} \partial_{\mu} \phi \partial_{\nu} \phi
+(\Box \phi)^2
\right),
\end{eqnarray}
(2)massive case
\begin{eqnarray}
S^{(0)}&=& -\frac{1}{2} \epsilon^{-4}\int d^4 x
\sqrt{-h} (4-\Delta) \phi^2, \nonumber\\
S^{(2)}&=& \frac{1}{4(\Delta-3)} \epsilon^{-2}
\int d^4 x \sqrt{-h}
\left( \frac{4-\Delta}{6} R \phi^2+h^{\mu \nu}
 \partial_{\mu} \phi \partial_{\nu} \phi \right),
\end{eqnarray}
where $R$, $R_{\mu \nu}$ and $\Box$ are defined in terms
of $h_{\mu \nu}$.

\end{document}